\documentclass{article}

\usepackage[numbers]{natbib}
\usepackage{rotating}
\usepackage{graphicx}
\usepackage{bm}

\newcommand{\threejm}[6]{\left(\begin{array}{ccc}#1 & #2 & #3 \\ #4 & #5 & #6 \end{array}\right)}

\begin{document}

\huge

\vspace{3cm}

\begin{center}
K-shell spectroscopy in hot plasmas: Stark effect, Breit interaction and QED corrections
\end{center}

\vspace{0.5cm}

\large

\begin{center}
Jean-Christophe Pain\footnote{jean-christophe.pain@cea.fr (corresponding author)}, Franck Gilleron and Maxime Comet
\end{center}

\vspace{0.2cm}

\normalsize

\begin{center}
CEA, DAM, DIF, F-91297 Arpajon, France
\end{center}

\vspace{0.2cm}

\large

\begin{center}
Dominique Gilles
\end{center}

\vspace{0.2cm}

\normalsize

\begin{center}
CEA, DSM, IRFU, F-91191 Gif-sur-Yvette, France
\end{center}

\vspace{0.5cm}


\begin{center}
{\bf Abstract}
\end{center}

The broadening of lines by Stark effect is widely used for inferring electron density and temperature in plasmas. Stark-effect calculations often rely on atomic data (transition rates, energy levels,...) not always exhaustive and/or valid only for isolated atoms. In this work, we first present a recent development in the detailed opacity code SCO-RCG for K-shell spectroscopy. The approach is adapted from the work of Gilles and Peyrusse. Neglecting non-diagonal terms in dipolar and collision operators, the line profile is expressed as a sum of Voigt functions associated to the Stark components. The formalism relies on the use of parabolic coordinates and the relativistic fine structure of Lyman lines is included by diagonalizing the hamiltonian matrix associated to quantum states having the same principal quantum number $n$. The SCO-RCG code enables one to investigate plasma environment effects, the impact of the microfield distribution, the decoupling between electron and ion temperatures and the role of satellite lines (such as Li-like $1sn\ell n'\ell' - 1s^2n\ell$, Be-like, etc.). Atomic structure calculations have reached levels of accuracy which require evaluation of Breit interaction and many-electron quantum electro-dynamics (QED) contributions. Although much work was done for QED effects (self-energy and vacuum polarization) in hydrogenic atoms, the case of an arbitrary number of electrons is more complicated. Since exact analytic solutions do not exist, a number of heuristic methods have been used to approximate the screening of additional electrons in the self-energy part. We compare different ways of including such effects in atomic-structure codes (Slater-Condon, Multi-Configuration Dirac-Fock, etc.).

\section{INTRODUCTION}

In hot dense plasmas encountered for instance in inertial confinement fusion (ICF), the line broadening resulting from Stark effect can be used to diagnose electronic temperature $T_e$, density $n_e$ and ionic temperature $T_i$. In previous versions of the SCO-RCG code \cite{Pain15a,Pain15b}, which was originally designed to perform detailed opacity calculations of complex (L-, M-, ... shell) spectra, the line shape resulted from the convolution of a Gaussian and a Lorentzian functions, leading to a so-called Voigt profile. The full-width at half-maximum (FWHM) of the Lorentzian was calculated from the electron broadening and natural radiative decay, and the FWHM of the Gaussian included the Doppler and ionic Stark (effect of the electric field of the neighboring ions) broadenings. In such a simplified model, the ionic Stark width was obtained from a semi-empirical approach proposed by Rozsnyai \cite{Rozsnyai77}. However, the use of Voigt profiles raises many questions, such as the asymptotic expansion of the wings or the fact that it precludes the accounting for asymmetry, and is definitely inadequate for K-shell spectra. The capability of the detailed opacity code SCO-RCG was recently extended to K-shell spectroscopy (hydrogen- and helium-like ions). The new developments address two main topics. The first one concerns Stark effect, which leads to a splitting of the lines, computed following an approach proposed by Gilles and Peyrusse \cite{Gilles15}. The second one is the inclusion of QED (self-energy and vacuum polarization) corrections and of the Breit interaction in the level and line energies.

\section{STARK EFFECT}

\section{Main assumptions and approximations}

In SCO-RCG, ions and electrons are treated respectively in the quasi-static and impact approximations and the line profile $\phi(\nu)$ is proportional to 

\begin{equation}
\frac{1}{\pi}\int\mathrm{Re}\left[\mathrm{Tr}\left\{\hat{\mathrm{d}}.\hat{\mathrm{X}}^{-1}\right\}\right]W(F)dF,
\end{equation}

\noindent where $\hat{\mathrm{X}}=2i\pi\left(\nu+\nu_1\right)-i\hat{\mathrm{H}}(F)/\hbar-\hat{\mathrm{\Lambda}}_c$, $\nu_1$ being the frequency of the lower state and $\hat{\mathrm{H}}(F)=\hat{\mathrm{H}}_0-\hat{\mathrm{d}}.F$ the Hamiltonian of the ion in the presence of an electric field $F$ following the normalized distribution $W(F)$. $\hat{\mathrm{H}}_0$ is the Hamiltonian without electric field while $\hat{\mathrm{d}}$ and $\hat{\mathrm{\Lambda}}_c$ represent respectively the dipole and collision operators. The trace (Tr) runs over the various states of the upper level. If $\Delta \nu_D$ is the Doppler width and $a_k$ the weight of the $k^{th}$ Stark component, neglecting non-diagonal terms in dipolar and collision operators, the line profile can be written as a sum of Voigt ($V$) functions (parametrized as in Ref. \cite{Humlicek79}):  

\begin{equation}
\phi(\nu)=\frac{1}{\sqrt{\pi}}\frac{1}{\Delta\nu_D}\int_0^{\infty}W(F)\left[\sum_ka_k(F)V(x_k,y_k)\right]dF\;\; ; \;\; x_k=\frac{\nu-\nu_0-c_k(F)}{\Delta\nu_D}\;\; ; \;\;  y_k=\frac{\langle k|\hat{\mathrm{\Lambda}}_c|k\rangle}{2\pi\Delta\nu_D},
\end{equation}

\noindent where $\nu_0$ is the frequency of the line without external field and

\begin{equation}
c_k(F)=\langle k|-\hat{\mathrm{d}}.F|k\rangle\;\;\; ; \;\;\; \hat{\mathrm{\Lambda}}_c=\frac{4\pi}{3}n_e\left(\frac{e}{\hbar}\right)^2\hat{\mathrm{d}}.\hat{\mathrm{d}}\left(\frac{2m}{\pi k_BT_e}\right)^{1/2}\ln\left(\frac{\lambda_{DH}Z}{n^2a_0}\right),
\end{equation}

\noindent $Z$ being the atomic number, $n$ the principal quantum number, $a_0$ the Bohr radius and

\begin{equation}
\lambda_{DH}=\sqrt{\frac{k_BT_e}{4\pi n_ee^2}}
\end{equation}

\noindent the Debye-H\"uckel length.

\section{Microfield distribution}

The microfield distribution function $W(F)$ is parametrized by ionic coupling $\Gamma=\left(Z^*e\right)^2/\left(r_{ws}k_BT_i\right)$ ($Z^*$ represents the average ionization) and electron degeneracy $\kappa=r_{ws}/\lambda_{\mathrm{TF}}$ constants, $r_{ws}$ being the Wigner-Seitz radius and $\lambda_{\mathrm{TF}}$ the Thomas-Fermi screening length, equal to 

\begin{equation}
\lambda_{\mathrm{TF}}^2=\lambda_{\mathrm{TF},0}^2.\frac{\left[12I_{1/2}\left(\frac{\mu}{k_BT_e}\right)\right]^{1/3}}{I_{-1/2}\left(\frac{\mu}{k_BT_e}\right)},
\end{equation}

\noindent where $\mu$ is the chemical potential and $\lambda_{\mathrm{TF},0}$ represents the Thomas-Fermi screening length at zero temperature, \emph{i.e.}

\begin{equation}
\lambda_{\mathrm{TF},0}^2=\frac{h^2}{16\pi^2me^2}\left(\frac{\pi}{3n_e}\right)^{1/3},
\end{equation}

\noindent where $n_e=Z^*n_i$, $n_i$ being the ionic density, and the Fermi integral reads

\begin{equation}
I_{n/2}(x)=\int_0^{\infty}\frac{y^{n/2}}{1+e^{y-x}}dy.
\end{equation}

We have the possibility to test many different microfield distributions. We generally use either the distribution computed from Monte Carlo simulations by Potekhin et al. \cite{Potekhin02}, or the one published by Laulan et al. \cite{Laulan08}, which is a combination of APEX (Adjustable Parameter EXponential) method \cite{Iglesias83} with a variational HNC (HyperNetted Chain) approach. The validity range of the obtained fitting formulas is:

\begin{equation}
\left\{\begin{array}{l}
10^{-3}\Gamma_m\leq\Gamma\leq\Gamma_m\;\;\;\; ;\;\;\;\;  0\leq\kappa\leq 4\\
10^{-1}\Gamma_m\leq\Gamma\leq\Gamma_m\;\;\;\; ;\;\;\;\;  4\leq\kappa\leq 5,
\end{array}\right.
\end{equation}

\noindent where $\Gamma_m$ is the coupling coefficient at melting temperature, computed as a fitting formula (see Ref. \cite{Hamaguchi97}):

\begin{equation}
\Gamma_m=\Gamma_m^{OCP}\times\frac{b(k^*)}{a(k^*)},
\end{equation}

\noindent where $\Gamma_m^{OCP}$ is the ionic coupling parameter in the One Component Plasma (OCP) model and

\begin{equation}
\left\{\begin{array}{l}
a(k^*)=1+1.0312\ln\left(k^*\right)+0.2674\left[\ln\left(k^*\right)\right]^2\\
b(k^*)=1+1.0200\ln\left(k^*\right)+0.4600\left[\ln\left(k^*\right)\right]^2+0.027\left[\ln\left(k^*\right)\right]^4, 
\end{array}\right.
\end{equation}

\noindent with the reduced parameter $k^*=1/(\kappa+1)$. Both microfield distributions of Refs. \cite{Potekhin02,Laulan08} are almost superimposed, but the formulas from Ref. \cite{Laulan08} have a wider range of validity, especialy as concerns the screening coefficient $\kappa$. 

\clearpage

\begin{figure}[h]
\begin{center}
  \includegraphics[width=200pt]{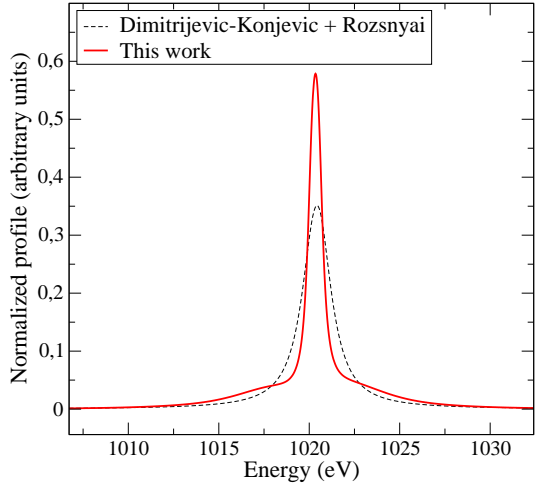}\vspace{3mm} 
  \caption{Ly$_{\alpha}$ line for a Ne plasma at $T_e$=$T_i$=200 eV and $\rho$=1 g/cm$^3$.}\label{figs1et2a}
\end{center}
\end{figure}

\begin{figure}[h]
\begin{center}
  \includegraphics[width=200pt]{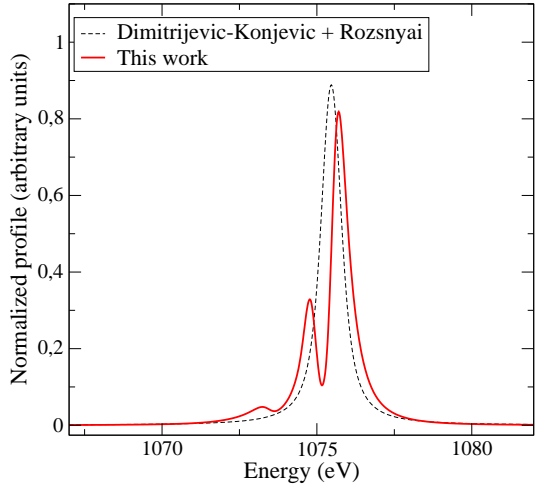}\vspace{3mm} 
  \caption{He$_{\beta}$ line for a Ne plasma at $T_e$=$T_i$=200 eV and $\rho$=0.01 g/cm$^3$.}\label{figs1et2b}
\end{center}
\end{figure}

\section{Hydrogen-like ions}

Stark effect for hydrogenic ions can be calculated in parabolic coordinates \cite{Bethe57} using the basis states $|nqm_{\ell}\rangle$, where $q=n_1-n_2$, $n_1$ and $n_2$ being the so-called parabolic quantum numbers, related by $n_1+n_2+|m_{\ell}|+1=n$, $-\ell\leq m_{\ell}\leq\ell$ being the magnetic orbital quantum number. The perturbation $\hat{\mathrm{d}}$ is diagonal in this basis and a 2$^{nd}$-order development gives

\begin{equation}
\langle nqm_{\ell}|-\hat{\mathrm{d}}.F|nqm_{\ell}\rangle=\frac{3}{2}\frac{ea_0}{Z}nqF-\frac{1}{16}\frac{e^2a_0^2}{(2Ry)}\left(\frac{n}{Z}\right)^4\left(17n^2-3q^2-9m_{\ell}^2+19\right)F^2.
\end{equation}

However, the fine-structure Hamiltonian $\hat{\mathrm{H}}_0$ is diagonal in the subset of states $|n\ell s j m_j\rangle$. In order to diagonalize the total Hamiltonian $\hat{\mathrm{H}}$ in such a basis, the Stark matrix element is

\begin{eqnarray}
\langle n\ell sjm_j|-\hat{\mathrm{d}}.F|n\ell'sj'm_j\rangle&=&\sum_{m_s=-1/2}^{1/2}\sum_{q=-\left(n-1-|m_{\ell}|\right)}^{n-1-|m_{\ell}|, 2}(-1)^{\ell+\ell'-1+3m_j-m_s-q-n}[\ell,\ell',j,j']^{1/2}\hspace{3cm}\nonumber\\
& &\times\threejm{\ell}{s}{j}{m_{\ell}}{m_s}{-m_j}\threejm{\ell'}{s}{j'}{m_{\ell}}{m_s}{-m_j}\nonumber\\
& &\times\threejm{\frac{n-1}{2}}{\frac{n-1}{2}}{\ell}{\frac{m_{\ell}-q}{2}}{\frac{m_{\ell}+q}{2}}{-m_{\ell}}\threejm{\frac{n-1}{2}}{\frac{n-1}{2}}{\ell'}{\frac{m_{\ell}-q}{2}}{\frac{m_{\ell}+q}{2}}{-m_{\ell}}\nonumber\\
& &\times\langle nqm_{\ell}|-\hat{\mathrm{d}}.F|nqm_{\ell}\rangle,
\end{eqnarray}

\noindent with $s=1/2$, $m_{\ell}+m_s=m_j$ and $[x]=2x+1$. Figure \ref{figs1et2a} displays a comparison between our previous semi-empirical modeling (Refs. \cite{Rozsnyai77,Dimitrijevic80}) and the present work in the case of Ne X Ly$_{\alpha}$ line at $T_e$=$T_i$=200 eV and $\rho$=1 g/cm$^3$. Figure \ref{figs1et2b} shows Ne IX He$_{\beta}$ profile at $T_e$=$T_i$=200 eV and $\rho$=0.01 g/cm$^3$.

\section{Helium-like ions}

We consider the transitions $1sn\ell$ $^1P-1s^2$, $n\ge 2$. For $n \geq 5$, the perturbation due to field $F$ is much larger than the separation between terms, the levels are quasi-hydrogenic and He lines are modeled as Ly-like lines with the substitution $Z\rightarrow Z-1$. For $n<5$, singlet-triplet mixing is neglected and the Hamiltonian $\hat{\mathrm{H}}_0-e(z_1+z_2)F$ is diagonalized in the sub-space of states $|1s; n\ell m_{\ell}; S\rangle$ with $S$=0 for singlet states and $S$=1 for triplet states. For He$_{\alpha}$, the resonance line ($1s2p~^1P - 1s^2$) requires the energies of terms $1s2s~^1S$ and $1s2p~^1P$ and the intercombination line ($1s2p~^3P - 1s^2$) the energies of terms $1s2s~^3S$ and $1s2p~^3P$.

\subsection{Interpretation of a ``buried-layer'' experiment on aluminum}

\begin{figure}[h]
\begin{center}
  \centerline{\includegraphics[width=230pt]{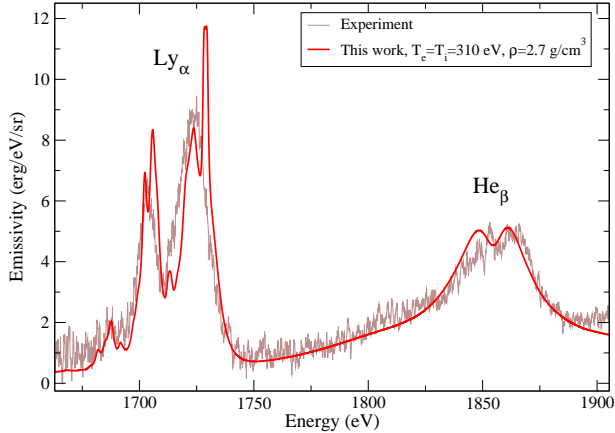}}\vspace{3mm} 
  \caption{Measured emission of aluminum ``buried layers'' heated by an ultra-short laser \cite{Dervieux15} (emissive volume: 400 $\mu m^2$ $\times$ 0.5 $\mu m$, duration: 3 ps) compared to SCO-RCG prediction.}\label{fig_der}
\end{center}
\end{figure}

Figure \ref{fig_der} shows our interpretation of the recently measured emission of aluminum micro-targets buried in plastic (``buried layers'') and heated by an ultra-short laser \cite{Dervieux15}. The agreement with the experiment is rather satisfactory, especially as the data are ``absolute values'' (no scaling was applied).

\vspace{10mm}

\begin{table}
\begin{center}
\begin{tabular}{cccccc}
 & $\mathbf{Breit,~QED}$ & $\mathbf{no}$ $\mathbf{Breit}$, $\mathbf{QED}$ & $\mathbf{no}$ $\mathbf{QED}$, $\mathbf{Breit}$ \\\hline
Resonance & 6699.84 & 6705.84 & 6703.50\\
Intercombination & 6666.25 & 6672.04 & 6670.11\\\hline
\end{tabular}\vspace{1cm}
\caption{Energies (in eV) of the resonance and intercombination lines He$_{\alpha}$ lines of iron with and without Breit interaction and QED corrections. The experimental values are 6700.01 eV (resonance) and 6668.11 eV (intercombination).}\label{tab1}
\end{center}
\end{table}
                                                            
\section{BREIT INTERACTION AND QED CORRECTIONS}

In its present version, the SCO-RCG code does not take into account Breit interaction and QED corrections. In order to study the importance of those contributions for our applications (see table \ref{tab1}), we used a Multi-Configuration Dirac-Fock (MCDF) code developed by J. Bruneau \cite{Bruneau83}. The Breit operator includes Coulomb repulsion, magnetic interaction and retardation in the electron-electron interaction due to finite value of the speed of light:

\begin{equation}
\hat{\mathrm{H}}_B=\frac{1}{r_{12}}-\frac{\bm{\alpha}_1.\bm{\alpha}_2}{r_{12}}\cos\left(\omega_{12}r_{12}\right)+\left(\bm{\alpha}.\bm{\nabla}\right)_1\left(\bm{\alpha}.\bm{\nabla}\right)_2\frac{\cos\left(\omega_{12}r_{12}-1\right)}{\omega_{12}^2r_{12}},
\end{equation}

\noindent where $\bm{\alpha}_i$ are the 4$\times$4 Dirac matrices, $\omega_{12}$ is the frequency of the exchange photon and the electron-electron interaction is expressed in the Coulomb gauge.

\begin{figure}[h]
\begin{center}
  \includegraphics[width=90pt]{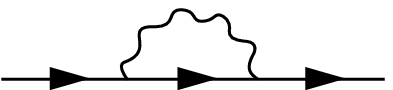}\hspace{1cm}
  \includegraphics[width=90pt]{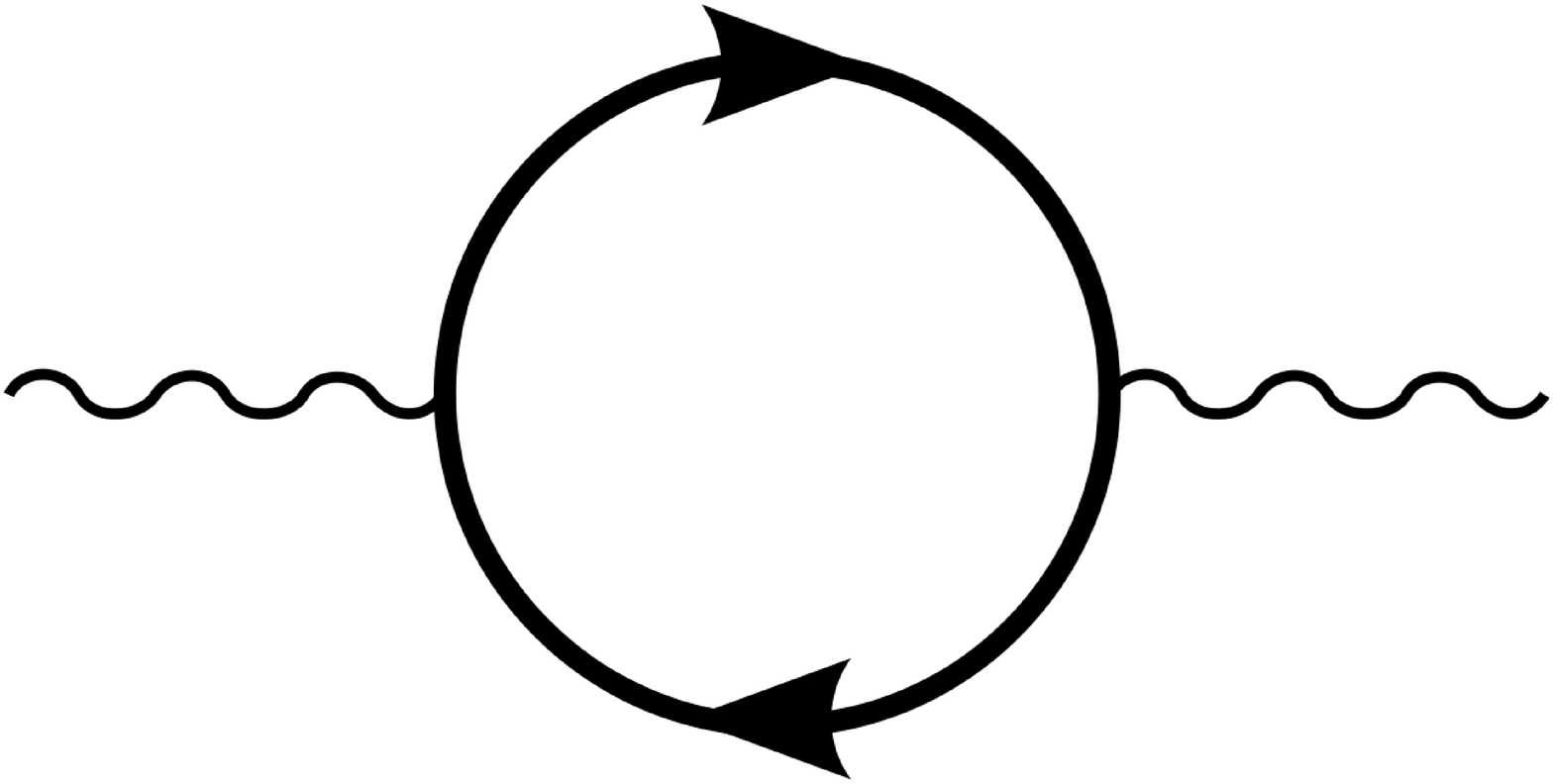}
  \caption{Feynman diagrams for self-energy (left side) and vacuum polarization (right side).}\label{fd}
\end{center}
\end{figure}

\begin{figure}[h]
\begin{center}
  \includegraphics[width=230pt]{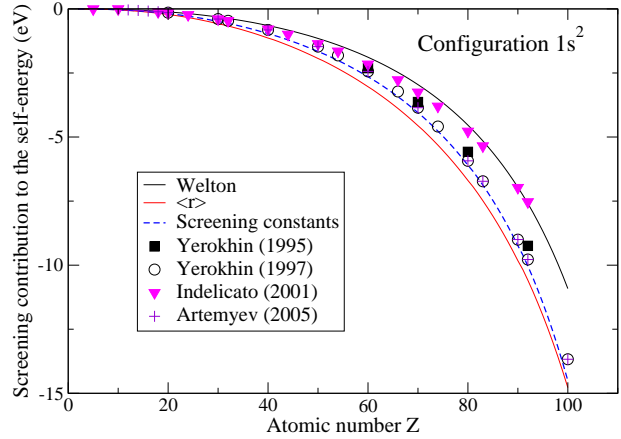}\vspace{3mm} 
  \caption{Self-energy screening contribution for $1s^2~J=0$ levels compared to the results of Refs. \cite{Yerokhin95,Yerokhin97,Indelicato01,Artemyev05}. }\label{1s2a}
\end{center}
\end{figure}

\begin{figure}[h]
\begin{center}
  \includegraphics[width=230pt]{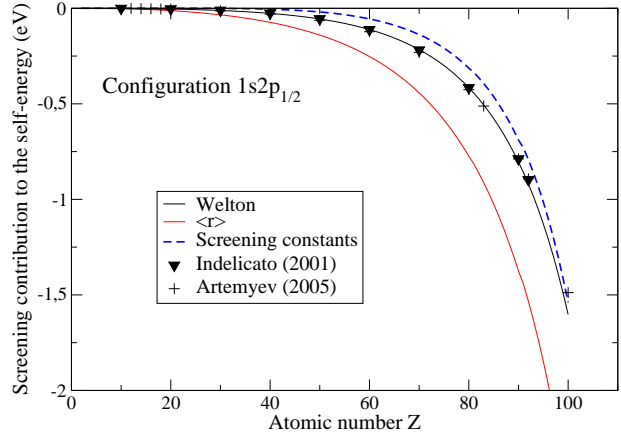}\vspace{3mm} 
  \caption{Self-energy screening contribution for $1s2p_{1/2}~J=1^-$ levels compared to the results of Refs. \cite{Indelicato01,Artemyev05}. }\label{1s2b}
\end{center}
\end{figure}

\begin{figure}[h]
\begin{center}
  \includegraphics[width=230pt]{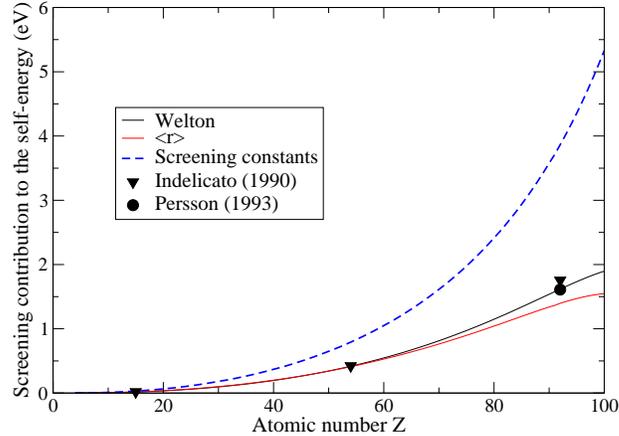}\vspace{3mm} 
  \caption{Self-energy screening contribution to the Lamb shift in the Li-like iso-electronic sequence compared to the values of Refs. \cite{Indelicato90,Persson93}.}\label{copa}
\end{center}
\end{figure}

\begin{figure}[h]
\begin{center}
  \includegraphics[width=230pt]{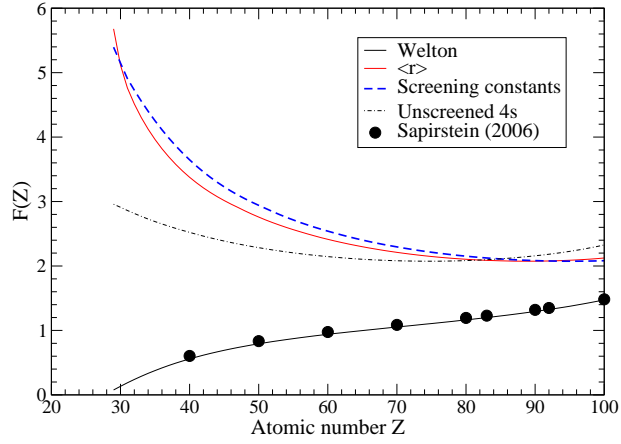}\vspace{3mm} 
  \caption{Function $F$ for 4s orbital in copper-like ions compared to the results of Ref. \cite{Sapirstein06}.}\label{copb}
\end{center}
\end{figure}

The so-called radiative corrections include vacuum polarization and self-energy (the Feynman diagrams are presented in Fig. \ref{fd}). Vacuum polarization is related to creation and annihilation of virtual electron-positron pairs in the field of the nucleus; it can be evaluated using effective potentials \cite{Uehling35}. The electromagnetic field of the electron can interact with the electron itself. In quantum field theory, this interaction corresponds to an electron emitting a virtual photon, which is then reabsorbed by the electron. The energy associated with this interaction is the self-energy of the electron (see tables \ref{tab2} and \ref{tab3}), responsible for the Lamb shift \cite{Lamb47}. The first self-energy calculations were carried out to first-order in $Z\alpha$ \cite{Bethe47}. In the early 1970s, Mohr proposed an atomic self-energy formulation within the bound-state Furry formalism in a suitable form for numerical evaluation:

\begin{equation}\label{foncf}
E_{n\ell j}^{\mathrm{SE}}\left(\alpha Z\right)=\frac{\left(\alpha Z\right)^4}{\pi n^3\alpha}F_{n\ell j}\left(\alpha Z\right),
\end{equation}

\noindent where $F$ is a slowly varying function of $\alpha Z$. For $s$ and $p$ orbitals, $F$ is evaluated using a development in powers of $(Z\alpha)$ and $\ln(Z\alpha)$  for $Z\leq 10$ \cite{Erickson77} and an interpolation in the tabulated values of Mohr \cite{Mohr74,Mohr75,Mohr82} for $Z>10$. For $n$=3 and 4 we take the fit published by Curtis \cite{Curtis85} and the results of Le Bigot et al. \cite{LeBigot01}. Calculation of many-electron radiative corrections is still one of the most difficult problems to deal with for high-precision level prediction. There have been no generalizations of the self-energy calculations to arbitrary $N-$electron systems. Without exact solutions, atomic-structure codes use an approximation to the self-energy which consists in evaluating the exact hydrogenic formulas of Mohr and successors for an effective charge $Z_{\mathrm{eff}}$ in order to account for screening and multiple-electron interactions. The screening contribution to the self-energy (as plotted in Figs. \ref{1s2a}, \ref{1s2b}, \ref{copa} and \ref{copb}) is defined as

\begin{equation}
E_{n\ell j}^{\mathrm{SE}}\left(\alpha Z_{\mathrm{eff}}\right)-E_{n\ell j}^{\mathrm{SE}}\left(\alpha Z\right)=\frac{\alpha^3}{\pi n^3}\left(Z_{\mathrm{eff}}^4F_{n\ell j}\left(\alpha Z_{\mathrm{eff}}\right)-Z^4F_{n\ell j}\left(\alpha Z\right)\right).
\end{equation} 

\begin{itemize}
\item The effective charge $Z_{\mathrm{eff}}$ can be determined from screening constants \cite{Lanzini15,DiRocco16}: the average charge of orbital $i$ is given by 
\end{itemize}

\begin{equation}
Z_{\mathrm{eff}}=Z-\left(\sum_{j<i}w_jf_{ji}+\sum_{j>i}w_jg_{ij}+(w_i-1)k_i\right)=Z-\sigma_i,
\end{equation}

\noindent where $w_i$ is the population of orbital $i$, and $f_{ji}$, $g_{ij}$ and $k_i$ the different screnning constants of the sub-shells, leading, after weighting by the electron populations, to the screening parameters $\sigma_i$.

\begin{itemize}
\item The effective charge $Z_{\mathrm{eff}}$ can be obtained from the average radius, solving $\langle r\rangle_{\mathrm{MCDF}}=\langle r\rangle_{\mathrm{hyd}}$ where
\end{itemize}

\begin{equation}
\langle r\rangle_{\mathrm{MCDF}}=\int_0^{\infty}\left(P_{n\ell j}^2(r)+Q_{n\ell j}^2(r)\right)rdr\;\;\;\;\mathrm{and}\;\;\;\;\langle r\rangle_{\mathrm{hyd}}=\frac{a_0}{2Z_{\mathrm{eff}}}\left[\left(3N^2-\kappa^2\right)\sqrt{1-\alpha^2Z^2/N^2}-\kappa\right]
\end{equation}

\noindent are respectively the average radius obtained from the MCDF wavefunctions ($P_{n\ell j}$ and $Q_{n\ell j}$ are respectively the small and large components) and the relativistic hydrogenic average radius of subshell $n\ell j$. One has $N=\sqrt{n^2-2n_r\left(|\kappa|-\gamma\right)}$ with $n_r=n-|\kappa|$ and $\gamma=\sqrt{\kappa^2-\alpha^2Z^2}$, where $\kappa=-\ell-1$ for $j=\ell+1/2$ and $\kappa=\ell$ for $j=\ell-1/2$.

\begin{figure}[h]
\begin{center}
  \includegraphics[width=230pt]{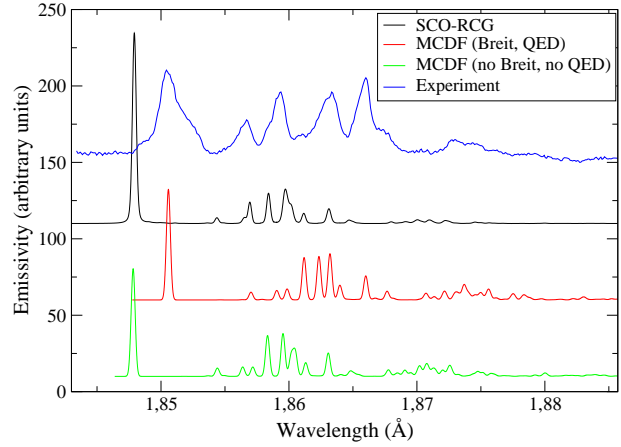}\vspace{3mm} 
  \caption{Comparison between emission spectrum of iron measured by Aglitskiy et al. \cite{Aglitskiy15} compared to several calculations at $T$=260 eV and $\rho$=0.01 g/cm$^3$: SCO-RCG \cite{Pain15a,Pain15b}, MCDF \cite{Bruneau83} and MCDF without Breit interaction and QED corrections.}\label{fig_nltea}
\end{center}
\end{figure}

\begin{figure}[h]
\begin{center}
  \includegraphics[width=237pt]{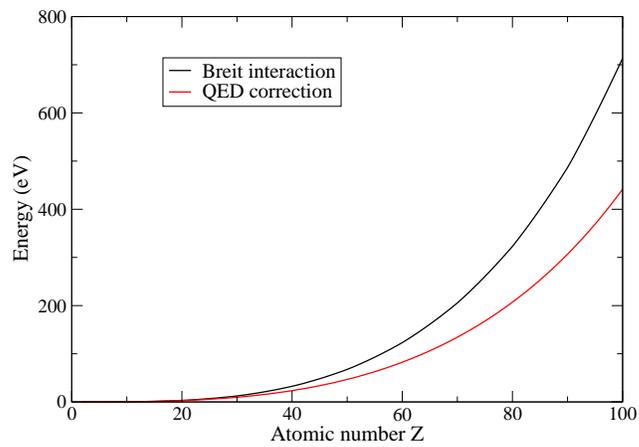}\vspace{3mm} 
  \caption{Contributions of Breit interaction and QED corrections to the energy of $1s^2$ as a function of atomic number $Z$. The MCDF wavefunctions are computed in the ``Slater transition state '' approximation \cite{Bruneau83}.}\label{fig_nlteb}
\end{center}
\end{figure}

\begin{table}
\begin{center}
\begin{tabular}{cccccc}
$\mathbf{Ion}$ & $\mathbf{Rodrigues}$ \cite{Rodrigues04} & $\mathbf{Curtis}$ \cite{Curtis85} & $Z_{\mathrm{eff}}$ $\mathbf{from}$ $\mathbf{screening}$ $\mathbf{constants}$ & $Z_{\mathrm{eff}}$ $\mathbf{from}$ $\langle r\rangle$ & $\mathbf{Welton}$ \\\hline
Li-like &  &  &  &  &\\
Z=55 & 116.90 & 112.12 & 113.73 & 113.96 & 144.89\\\hline
Z=95 & 886.48 & 923.46 & 883.28 & 882.51 & 887.69\\\hline
Na-like &  &  &  &  &\\
Z=55 & 131.16 & 131.32 & 123.51 & 122.89 & 125.15\\\hline
Z=95 & 1052.00 & 1046.85 & 1006.44 & 993.85 & 1024.40\\\hline
\end{tabular}
\caption{Self-energy (in eV) for Li-like and Na-like Cs and Am computed with our MCDF code \cite{Bruneau83}.}\label{tab2}
\end{center}
\end{table}

\begin{table}
\begin{center}
\begin{tabular}{cccc}
$\mathbf{Orbital}$ & $\mathbf{Self-energy}$ & $\mathbf{Vacuum}$ $\mathbf{polarization}$ & $\mathbf{Total}$ $\mathbf{QED}$ \\\hline
$1s_{1/2}$ & 357.566 & -93.824 & 263.742 \\
$2s_{1/2}$ & 66.073 & -16.517 & 49.556 \\
$2p_{1/2}$ & 9.608 & -0.127 & 9.481 \\\hline
\end{tabular}\vspace{1cm}
\caption{Self-energy, vacuum polarization and total QED contributions (in eV) for orbitals $1s_{1/2}$, $2s_{1/2}$ and $2p_{1/2}$ in U$^{91+}$ computed with our MCDF code \cite{Bruneau83}.}\label{tab3}
\end{center}
\end{table}

\clearpage

\noindent In Welton's picture of the Lamb shift \cite{Welton48,Indelicato87}, the self-energy is due to perturbations of the classical trajectory of the electron by fluctuations of the vacuum's electromagnetic field. These fluctuations cause the electron to probe the potential at a displaced point $V_N\left(\mathbf{r}+\delta\mathbf{r}\right)$ rather than $V_N\left(\mathbf{r}\right)$. This yields a perturbing potential

\begin{eqnarray}
\delta V_N&=&\langle V_N\left(\mathbf{r}+\delta\mathbf{r}\right)-V_N\left(\mathbf{r}\right)\rangle_{\mathrm{vacuum}}\nonumber\\
&\approx &\langle \bm{\nabla}V_N\left(\mathbf{r}\right).\delta\mathbf{r}+\Delta V_N\left(\mathbf{r}\right)\left(\delta\mathbf{r}\right)^2+\cdots\rangle_{\mathrm{vacuum}}\nonumber\\
&\approx &\langle\Delta V_N\left(\mathbf{r}\right)\left(\delta\mathbf{r}\right)^2\rangle_{\mathrm{vacuum}},
\end{eqnarray}

\noindent where the first term on the right-hand side of the second line vanishes because the vacuum fields average to zero. The non-vanishing second term must be renormalized, and gives the hydrogenic formula (\ref{foncf}). Welton therefore argues that, at least for $s$ orbitals, it is more relevant to use the \emph{ansatz} 

\begin{equation}
E_{n\ell j}^{SE}\left(\mathrm{MCDF}\right)=\frac{\langle n\ell j|\Delta V_N(r)|n\ell j\rangle_{\mathrm{MCDF}}}{\langle n\ell j|\Delta V_N(r)|n\ell j\rangle_{\mathrm{hyd}}}E_{n\ell j}^{SE}\left(\mathrm{hyd}\right)\;\;\;\;\mathrm{where}\;\;\;\;\Delta V_N(r)\propto\rho_N(r)=\frac{\rho_0}{1+\exp\left[\left(r-R_N\right)/t\right]},
\end{equation}

\noindent $\rho_0$ being obtained from

\begin{equation}
Z=\int_0^{\infty}4\pi r^2\rho_N(r)dr,\;\;\;\; \mathrm{i.e.}\;\;\;\;\rho_0=\frac{3Z}{4\pi R_N^3\mathcal{N}}\;\;\;\;\mathrm{with}\;\;\;\;\mathcal{N}=1+\frac{\pi^2t^2}{R_N^2}+6\frac{t^3}{R_N^3}\sum_{n=1}^{\infty}\frac{(-1)^{n-1}}{n^3}e^{-\frac{nR_N}{t}}.
\end{equation}

\noindent This two-parameter model has a uniform core with a ``skin'' in which the density falls from 90 \% to 10 \% of its central value in a short distance. We take the thickness parameter $t\approx$ 1.0393 10$^{-5}$ at. u. and the RMS radius of the nuclear charge distribution $R_N\approx$ 2.2677 10$^{-5}$ $A^{1/3}$ at. u., $A$ being the atomic mass (in g.). 

Figure \ref{1s2a} shows that for $1s^2~J=0$ the ``average radius'' and ``screening constants'' approaches yield values similar to eachother and closer to the Artemyev many-body-perturbation-theory reference calculations than Welton's picture. In other cases, screening constants seem to be more relevant than $\langle r\rangle$, the best approach being probably Welton's picture of the Lamb shift \cite{Lowe13}, as can be seen on Figs. \ref{1s2b}, \ref{copa} and \ref{copb}.

Figure \ref{fig_nltea} displays an experimental spectrum of iron measured on the NRL KrF Nike laser facility, capable of delivering several kilojoules of ultraviolet light ($\lambda$=248 nm) on a target within a few nanoseconds which is sufficient to produce high-Z ions with multi-keV ionization potentials. As such this system is a powerful platform to benchmark high-energy-density plasma diagnostics and relevant atomic-physics simulations. For this purpose an imaging spectrometer using a spherically curved crystal provides high-resolution spectra within a narrow variable spectral band. The experimental spectrum of Fig. \ref{fig_nltea} is clearly out of local thermodynamic equilibrium, and we do not intend to reproduce the relative intensities of the different lines. We only care of the position of the lines and the values of QED corrections and Breit interaction for $1s^2$ in that case are plotted on Fig. \ref{fig_nlteb}. We can see that the SCO-RCG calculation (which does not include Breit and QED corrections) does not reproduce the experimental line energies, whereas MCDF (which includes Breit and QED corrections) clearly does. In addition, when we cancel the Breit and QED corrections in MCDF, we recover the line energies predicted by SCO-RCG, which means that Breit and QED corrections have to be included in SCO-RCG and that, in that case, the impacts of exchange-correlation modeling and density effects are not so important.   

\section{CONCLUSION}

The SCO-RCG code was originally designed to perform detailed opacity calculations. Therefore, in its previous versions, due to the huge number of lines included in the computations, the line shapes were simply modeled by Voigt profiles. Such an approach is usually sufficient for L- and M-shell opacities, where the lines are so numerous that they overlap at least partially leading to complicated but mostly unresolved structures. However, the use of Voigt profiles raises many questions, such as the truncation of the wings, and is undoubtedly irrelevant for K-shell spectra. We have presented recent developments in the SCO-RCG code concerning K-shell spectroscopy. We first replaced the Voigt functions by real Stark profiles and included the contribution of QED (self-energy and vacuum polarization) and of the Breit interaction in the line energies. In the future, we plan to investigate the importance of autoionizing states $1s 2\ell 2\ell'$ and $1s2\ell 3\ell'$ of He$_{\beta}$ (in the present work we only took into account $2\ell 2\ell'$) and to include the line $1s3d~^1D_2$ - $1s^2~^1S_0$ induced by the field (mixing states $1s3d~^1D_2$ and $1s3p~^1P_1$) as well as the lines $1s3d~^3D_2$ - $1s^2~^1S_0$ and $1s3s~^3S_1$ - $1s^2~^1S_0$. We also started to study the Stark-Zeeman splitting. It is important to mention that the impact of Breit interaction and QED corrections, although very small, can play a significant role in the interpretation of hot-plasma K-shell emission spectra.

\end{document}